\documentclass[twocolumn,floatfix,amsmath,amssymb,aps,pre]{revtex4-1}

\usepackage{amssymb}         
\usepackage{bm}
\usepackage{xcolor}          
\usepackage{graphicx}

\begin{document}

\title{Elastic turbulence in two-dimensional cross-slot viscoelastic flows}

\author{D. O. Canossi} 
\author{G. Mompean} 
\author{S. Berti}

\affiliation{Univ. Lille, ULR 7512 - Unit\'e de M\'ecanique de Lille Joseph Boussinesq (UML), F-59000 Lille, France}


\begin{abstract}
We report evidence of irregular unsteady flow of two-dimensional polymer solutions in the absence of inertia in cross-slot geometry using numerical simulations of Oldroyd-B model. By exploring the transition to time-dependent flow versus both the fluid elasticity and the polymer concentration, we find periodic behaviour close to the instability threshold and more complex flows at larger elasticity, in agreement with experimental findings. For high enough elasticity we obtain dynamics pointing to elastic turbulence, with temporal spectra of velocity fluctuations showing a power-law 
decay, of exponent in between $-3$ and $-2$,  
and probability density functions of velocity fluctuations that 
weakly deviate from Gaussianity while high non-Gaussian tails characterise those of local accelerations.
\end{abstract}

\maketitle

\section*{Introduction}
\label{sec:intro}
The rheological behaviour of viscoelastic flows at vanishingly small inertia can be related 
to strongly non-linear phenomena and includes an association of viscous and elastic effects, 
with the latter being typically due to the presence of flexible long-chain polymers in the solution. 
The elasticity of the flow can give rise to complex dynamics that are relevant for both fundamental 
studies and industrial applications, as \textit{e.g.} efficient mixing and heat transfer in microdevices~\cite{Traore-15}, 
or painting and coating processes~\cite{Larson-92,Shaqfeh-96,Li-12}.

The purely elastic instabilities marking the transitions between different flow regimes 
have been documented in a variety of geometrical configurations~\cite{Muller-89,Shaqfeh-96,Groisman-00,Berti-08}, 
including complex ones, such as the abrupt axisymmetric contraction~\cite{McKinley-91} 
and the lid-driven cavity~\cite{Pakdel-96}. The cross-slot setup, made of two perpendicularly intersecting channels 
with two inlets and two outlets, is, in this sense, no exception. Due to its relevance for mixing and rheology, 
it has been the subject of extensive studies. 
Indeed, experimental~\cite{Scrivener-79,Arratia-06,Sousa-15}, theoretical~\cite{Harlen-90,Becherer-08} 
and numerical~\cite{Remmelgas-99,Poole-07,Rocha-09} investigations have reported about the existence of instabilities 
solely driven by elasticity in this setup. It is now known that low-Reynolds-number polymeric flows in this geometry 
can display two types of instabilities: a first one, corresponding to a supercritical pitchfork bifurcation, 
to steady asymmetric flow~\cite{Poole-07,Rocha-09}, and a second one leading to unsteady oscillatory 
behaviour~\cite{Arratia-06,Poole-07,Haward-12}. Concerning the latter, it is interesting to recall that it has been provided 
numerical evidence, in a two-dimensional (2D) flow, that it occurs via a supercritical Hopf bifurcation~\cite{Xi-09}; 
a mechanism relying on the role of stress gradients and the existence of a stagnation point at the centre of the setup 
was also proposed~\cite{Xi-09}. 

Above a critical Weissenberg number ($Wi$), meaning for elasticity larger than a threshold, 
purely elastic instabilities can lead to the appearance of disordered
flows corresponding to the dynamical regime known as elastic turbulence~\cite{Vinogradov-65,Groisman-00}. 
As shown in the seminal work~\cite{Groisman-00}, 
where a swirling flow between counter-rotating parallel disks was considered, 
and in subsequent ones also employing different geometries~\cite{Groisman-01,Groisman-04}, such flows are reminiscent 
of the turbulent ones occurring in Newtonian fluids. In particular, they are characterised by a whole range of active scales, 
irregular temporal behaviour, growth of flow resistance and enhanced mixing properties~\cite{Groisman-01}. 
Interestingly, however, the spectrum of velocity fluctuations displays power-law behaviours, in both the temporal 
($E(f) \sim f^{-\delta}$) and spatial ($E(k) \sim k^{-\delta}$) domains, with an exponent 
(in absolute value) $\delta \approx 3.5>3$, 
corresponding to a smooth flow essentially dominated by the largest spatial scales. 
It is worth to remark that such experimental findings are supported by theoretical predictions based on a simplified 
uniaxial model of viscoelastic fluid dynamics in the absence of walls and in homogeneous isotropic 
conditions~\cite{Fouxon-03}. 
At the same time, it was recently pointed out in~\cite{Gupta-19} that numerical simulations based on standard 
constitutive models may be dramatically affected by the polymer-stress diffusivity typically 
added to the evolution equations to ensure numerical stability, and that this particularly applies to flows 
characterised by regions of pure strain. Notably, using a cellular forcing in two dimensions, it was shown that 
kinetic energy spectra are considerably flatter in the absence of artificial polymeric diffusion and 
scale as $k^{-2.5}$~\cite{Gupta-19}. 

The elasticity-driven transition to turbulent-like states was experimentally investigated 
in cross-slot devices of different aspect ratio (vertical size over channel width), for more and less concentrated 
polymer solutions~\cite{Sousa-18}. Independently of the aspect ratio, it was found that the more concentrated solution 
undergoes a transition to unsteady flows that become progressively more irregular when the Weissenberg number is increased. 
The power spectra of velocity fluctuations, obtained from single-point time series of the streamwise component 
measured in the outlet channel at midway from the lateral walls 
(both in the horizontal and vertical directions), were characterised by the presence of marked peaks 
(a fundamental frequency plus some harmonics), and by a power-law behaviour of exponent 
smaller than $-3$, 
at small and large $Wi$ values, respectively. In particular, for the smaller aspect ratio, continuous spectra 
and features typical of elastic turbulence were observed when $Wi \gtrsim 25$. 
For the more dilute solution, although the phenomenology of the transitions was similar, the chaotic flow observed 
at high $Wi$ did not show similar spectral properties. 

In this letter we explore the unsteady viscoelastic flow regime occurring in a 2D cross-slot geometry at high elasticities 
and vanishing Reynolds number ($Re$) by means of extensive numerical simulations, for different polymer concentrations. 
For this purpose we adopt Oldroyd-B model, \textit{i.e.} the simplest possible one, to describe the dynamics of the viscoelastic fluid. 
As in~\cite{Buel-18}, where elastic turbulence was simulated in a 2D Taylor-Couette system, we 
integrate the model evolution equations using the open-source code 
OpenFOAM\textsuperscript{\textregistered}~\cite{Weller-98,Pimenta-17}, 
which allows control of the numerical instabilities associated with large Weissenberg numbers~\cite{Sureshkumar-95}. 
We provide numerical evidence of the emergence of turbulent-like features   
for quite concentrated solutions when $Wi$ is large enough. 
We analyse the transition to irregular dynamics and we characterise the statistical properties of the high-$Wi$ flows, 
discussing the similarities and differences with experimental results. 

\section*{Model and methods}
\label{sec:methods}

\begin{figure}[t]
  \centering
  \includegraphics[width=0.38\textwidth]{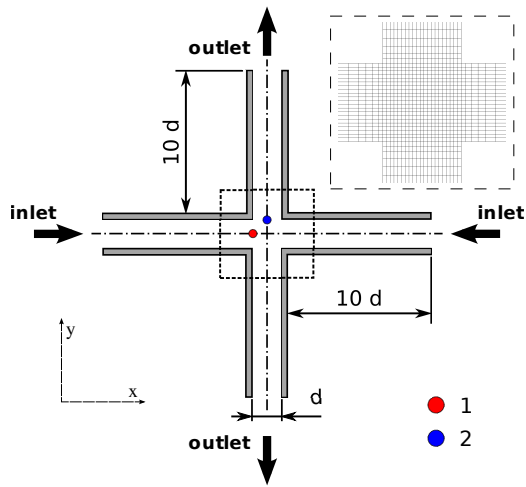}
  \caption{(Colour online) Schematic of the cross-slot geometry. The dotted square is the area where the analyses 
  were conducted, with the two dots indicating the positions where time series were recorded: 
  probe 1 (entrance, red), probe 2 (exit, blue). 
  Inset: zoom of the central area and typical mesh refining towards the centre of the setup; 
  note that simulations were performed with at least twice finer meshes.
  }
  \label{fig:f1}
\end{figure}

We consider an isothermal, incompressible, inertialess, 2D viscoelastic fluid flow in a cross-slot geometry. The latter consists of two perpendicular and bisecting channels of identical width $d$, with opposing inlets (here, along the $x$ direction) and outlets (along the $y$ direction), as shown schematically in fig.~\ref{fig:f1}.
The velocity field $\bm{u}(\bm{x},t)=(u_x(\bm{x},t), u_y(\bm{x},t))$ at position $\bm{x}$ and time $t$ evolves according to the momentum conservation equation
\begin{equation}
\label{eq:momentum}
\rho \left[ \frac{\partial \bm{u}}{\partial t} + \left( \bm{u} \cdot \bm{\nabla} \right) \bm{u} \right] = 
\bm{\nabla} \cdot \bm{T} - \bm{\nabla} p
\end{equation}
and the incompressibility condition $\bm{\nabla} \cdot \bm{u} = 0$, where $\bm{T}$ is the total (viscous plus elastic) stress tensor, $p$ the pressure and $\rho$ the density.

In the framework of Oldroyd-B model~\cite{Oldroyd-50,Bird-87}, the stress tensor $\bm{T}$ is the sum of a viscous component $\bm{\sigma} = \eta_s \, \dot{\bm{\gamma}}$, with $\eta_s$ the zero-shear dynamic viscosity of the solvent and $\dot{\bm{\gamma}}=\bm{\nabla} \bm{u} + \left(\bm{\nabla} \bm{u} \right)^T$ the strain-rate tensor, and an elastic one $\bm{\tau}$ due to polymers. The constitutive equation for the extra-stress tensor $\bm{\tau}$ reads:

\begin{equation}
\label{eq:oldroyd-b}
\bm{\tau} + \lambda \left[\frac{\partial \bm{\tau}}{\partial t} + \bm{\nabla} \cdot \left(\bm{u} \bm{\tau} \right) - \left( \bm{\nabla} \bm{u} \right)^T \cdot \bm{\tau} - \bm{\tau} \cdot \bm{\nabla} \bm{u} \right] = \eta_p \, \dot{\bm{\gamma}} \, ,
\end{equation}
where $\lambda$ represents the largest polymer relaxation time and $\eta_p$ the polymer contribution to viscosity. An important parameter is the viscosity ratio $\beta=\eta_s/(\eta_s+\eta_p)$, which is inversely proportional to the polymer concentration. Let us remark that in the limit $\beta \to 0$, one recovers the upper-convected Maxwell (UCM) model~\cite{Bird-87}, accounting for the dynamics of very concentrated solutions. At fixed $\beta$, the control parameters of the dynamics specified by eqs.~(\ref{eq:momentum}) and (\ref{eq:oldroyd-b}) are the Reynolds 
$Re=\rho U_b d/(\eta_s+\eta_p)$  
and Weissenberg $Wi=\lambda U_b/d$ numbers, where $U_b$ is the (uniform) velocity at the inlet.

In spite of important limitations, such as the infinite extensibility of polymers - and the consequent 
unbounded nature of extensional viscosity at strain rates $\ge 1/(2\lambda)$ - or the absence 
of shear-dependent viscosity, Oldroyd-B model corresponds to the simplest differential constitutive equation 
for viscoelastic fluids, and it exhibits normal stress differences. Furthermore, it has been successfully employed 
to numerically reproduce the basic phenomenology of elastic turbulence in different 2D 
configurations~\cite{Berti-08,Plan-17,Buel-18}.

\subsection*{Numerical simulations}
\label{ssec:num-sim}

Equations (\ref{eq:momentum}) and (\ref{eq:oldroyd-b}) are integrated by means of the open-source 
numerical solver \textsc{rheoTool}\textsuperscript{\textregistered}, which was developed in the framework 
of the OpenFOAM\textsuperscript{\textregistered} simulation code~\cite{Pimenta-17}. This solver is based 
on a finite-volume discretisation and makes use of the log-conformation technique~\cite{Fattal-04} 
to control the numerical instabilities appearing at high $Wi$ values. We remark that no 
polymer-stress diffusion is included. 

The cross-slot configuration has recently been proposed as a benchmark problem~\cite{Cruz.F-14}, 
for its geometrical characteristics and the existence of the instability leading to asymmetric flow 
at appropriate $\beta$, $Re$ and $Wi$ values. 
Similarly to the reference studies with this setup, here we set a length to width ratio of $10:1$ 
for each of the four ``arms'', which was previously shown to be enough to ensure a fully developed flow 
away from the inlet in a channel~\cite{Durst-05}. 
The global mesh adopted 
for the numerical integration is composed of four blocks, each of which corresponds to an arm, 
with increased density of grid points when approaching the centre of the system, plus a central square block 
with the smallest (uniform) grid size. The results presented in the following were obtained 
with a total of $12801$ computational cells, corresponding to $51 \times 51$ cells and a minimal grid spacing 
$\Delta x_{min} = \Delta y_{min} \approx 0.02 \, d$ in the central region. The mesh refinement towards the centre 
in each arm is realised via a geometric progression relation with a stretching factor $f_s = 0.931$. 
In order to verify the robustness of our results, some calculations, and particularly those related 
to the instability thresholds, were repeated with a mesh twice as refined. 
The results were qualitatively independent of the mesh size and only slight differences in the values 
of the critical parameters were found. The general phenomenology in the developed regime is also found 
to hold similar using the more refined mesh.

A uniform velocity profile of amplitude $U_b$ is applied at both inlets, where a homogeneous Neumann (zero gradient) 
boundary condition is specified for the pressure field, whereas polymeric extra-stresses are set to zero. 
At the outlets, a homogeneous Dirichlet (zero value) boundary condition is imposed for pressure, as well as 
zero-gradient ones for velocity and extra-stress fields. At the walls, no-slip conditions ($\bm{u}=0$) are applied 
to the velocity field and a linear extrapolation technique is adopted for the extra-stresses~\cite{Pimenta-17}. 
The velocity and stress initial condition corresponds to no flow.

The Weissenberg number was varied by changing the polymer relaxation time $\lambda$ only; the polymer concentration 
was set by choosing $\eta_s$ and $\eta_p$ such that their sum is constant. 
The Reynolds number, accounting for the relative strength of the non-linear inertial term 
to the viscous one in eq.~(\ref{eq:momentum}), was kept fixed at $Re=0$ by neglecting the term 
$(\bm{u} \cdot \bm{\nabla}) \bm{u}$ in eq.~(\ref{eq:momentum})~\cite{Rocha-09}, but we checked that including 
the latter (and setting $Re=0.1$) did not strongly affect the results on the instability 
critical parameters. 
Further, the dynamics appear not to be very sensitive to the presence of the term $\rho \partial_t \bm{u}$ 
in eq.~(\ref{eq:momentum}).

\section*{Results}
\label{sec:results}

\begin{figure}[bt]
  \centering
  \includegraphics[scale=0.45]{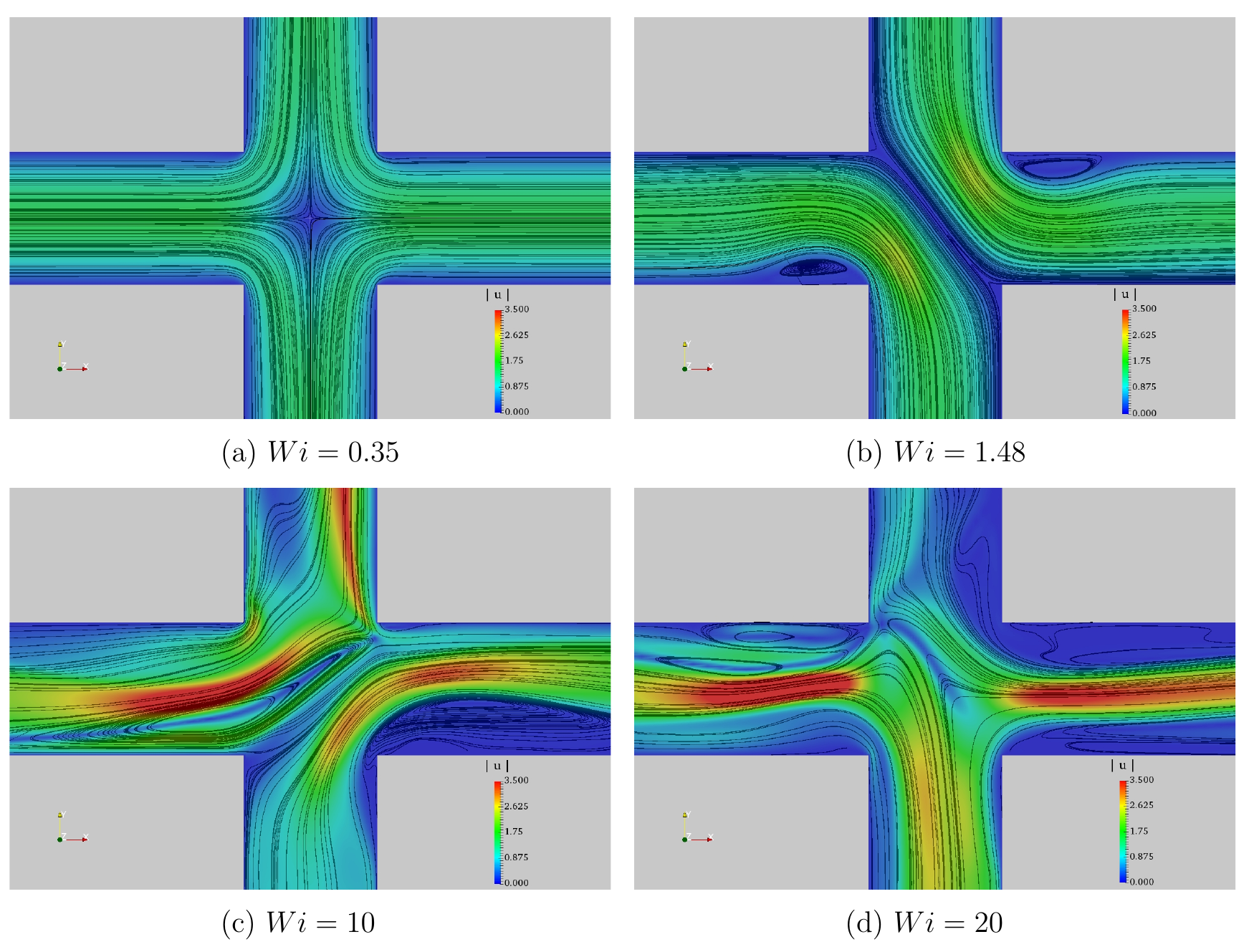}
  \caption{(Colour online) Snapshots of the magnitude of the velocity field (colour) and flow streamlines 
  (black lines) for $\beta=1/9$ and $Re=0$. Increasing $Wi$, different regimes are observed: 
  steady symmetric (a), steady asymmetric (b), unsteady disordered flow (c, d).
} 
\label{fig:f2}
\end{figure}

When increasing the elasticity of the solution, while keeping $\beta$ fixed, in our numerical integrations, 
we observe a destabilisation of the flow, in agreement with previous studies~\cite{Arratia-06,Poole-07}. 
The sequence of flow states that are selected depends on the polymer concentration, however, and here we provide 
a full picture of the stability portrait of the system as a function of both $\beta$ and $Wi$. 
Let us preliminarily remark that below the onset of purely elastic instabilities the flow coming from each of the inlets 
splits into two streams of equal flow rate, a symmetric state, at the outlets (see fig.~\ref{fig:f2}a). 
For concentrated solutions ($\beta \lesssim 0.56$), 
the flow first transitions to a steady asymmetric state (fig.~\ref{fig:f2}b, where $\beta=1/9$). 
By measuring the degree of asymmetry, expressed in terms of the excess flow rate in a stream, as a function of $Wi$ 
we verified (results not shown) that this transition is a supercritical pitchfork bifurcation. 
Our values of the critical Weissenberg number are in good agreement with those reported 
in previous benchmark studies~\cite{Cruz.F-14} (relative difference of less than $0.05$) 
both for $\beta=1/9$ and the UCM case $\beta=0$. 
In this range of low $\beta$ values, a second instability manifests when $Wi$ is further increased 
beyond a second threshold value close to $1$, leading to time-dependent behaviour in the form 
of regular oscillations of the asymmetric flow pattern (which stays similar to that of fig.~\ref{fig:f2}b). 
The situation changes for more diluted solutions (\textit{i.e.} when $\beta \gtrsim 0.56$). Indeed, in this case, 
the steady asymmetric flow regime does not set in and a direct change from steady symmetric to unsteady flow is observed. 
Remarkably, the same qualitative phenomenology is also found in experiments in micro-scale devices~\cite{Sousa-15}. 
In the time-dependent regime, and particularly for low $\beta$, an increase of $Wi$ eventually gives rise 
to spatially and temporally more complex flows akin to elastic turbulence ones. Two illustrative examples at fixed time 
are shown in fig.~\ref{fig:f2}c,d for $\beta=1/9$ and two different values of $Wi$. 

\begin{figure}[htbp]
  \centering
  \includegraphics[width=0.38\textwidth]{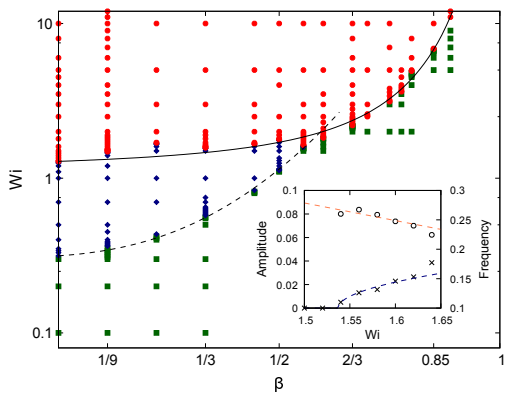}
  \caption{(Colour online) Stability diagram in the $(\beta,Wi)$ plane at $Re=0$. 
   The green squares, blue diamonds and red dots respectively correspond to steady symmetric, 
   steady asymmetric and unsteady flow. 
   The dashed ($Wi_c^{(\mathrm{I})}$) and continuous ($Wi_c^{(\mathrm{II})}$) lines are fits using eq.~(\ref{eq:Wic}); 
   here $a_0^{(\mathrm{I})} \simeq 2.75$, $a_{-1}^{(\mathrm{I})} \simeq -3.94$, 
   $a_0^{(\mathrm{II})} \simeq 0.85$, $a_{-1}^{(\mathrm{II})} \simeq 0.05$. 
   Inset: amplitude and frequency of $|\bm{u}(\bm{x}_*^{(2)},t)|$ vs $Wi$ at the onset of unsteady flow, 
   for 
   $\beta=1/9$.
   }
  \label{fig:f3}
\end{figure}
The complete stability portrait, obtained by spanning the ($\beta,Wi$) plane with a large number of simulations, 
is shown in fig.~\ref{fig:f3}, where the 
different point types correspond to the different dynamical regimes observed; 
here we only show a limited subset of the results from the simulations performed.  
By measuring the amplitude and frequency of the time series of $|\bm{u}(\bm{x}_*^{(2)},t)|$ at the fixed location $\bm{x}_*^{(2)}$ 
(corresponding to point $2$ in fig.~\ref{fig:f1}) for $Wi$ close to the onset of the unsteady regime 
and for different concentration values, we could assess that the second instability is a supercritical Hopf bifurcation 
(see inset of fig.~\ref{fig:f3} for $\beta=1/9$), as also suggested by~\cite{Xi-09} using a FENE-P model 
at non-zero $Re$ and large $\beta$. Indeed, the velocity signal displays a growth of its amplitude that is 
fairly well described by $(Wi-Wi_c^{(\mathrm{II})})^{1/2}$, with $Wi_c^{(\mathrm{II})}$ the critical Weissenberg number, 
and an approximately linear decrease of its frequency with $Wi$. 
For both the first and the second instability, the critical Weissenberg number, $Wi_c^{(\mathrm{I})}$ and $Wi_c^{(\mathrm{II})}$ respectively, 
grows with growing $\beta$, which is reasonable since increasing $\beta$ corresponds to decreasing polymer concentration. 
The faster growth of $Wi_c^{(\mathrm{I})}(\beta)$ causes the shrinking of the region of steady asymmetric flow. 
Determining the functional dependencies $Wi_c^{(i)}(\beta)$ (with $i=\mathrm{I}, \mathrm{II}$) from stability analysis is not an easy task, 
due to the formation of a birefringent strand and a diverging base state associated with the infinite extensibility 
of polymers~\cite{Becherer-08}. 
Since here we are mainly interested in characterising the boundaries, in the $(\beta,Wi)$ plane, 
of the regions where elastic turbulence could be excited, we proceed heuristically, especially focusing on $Wi_c^{(\mathrm{II})}(\beta)$. 
In order to account for non-zero $\beta$ effects, we conjecture that $Wi_c^{(\mathrm{II})}(\beta)=Wi_c^{(\mathrm{II})}(0)f(\beta)$, where 
$f(\beta)$ is a positive analytic function, except for $\beta \to 1$ where a divergence is expected, 
since the fluid becomes Newtonian and no purely elastic instability should occur; clearly $f(0)=1$. 
Our numerical results suggest that the data are compatible with a Laurent expansion at second order around the point 
$\beta=1$. Somehow more surprisingly, we find that the same functional shape can also be used to fit 
the $Wi_c^{(\mathrm{I})}(\beta)$ data, indicating that: 
\begin{equation}
\label{eq:Wic}
Wi_c^{(i)} = Wi_c^{(i)}(0)\left[ a_0^{(i)} + \frac{a_{-1}^{(i)}}{1-\beta} + \frac{a_{-2}^{(i)}}{(1-\beta)^2} \right] \, ,
\end{equation}
where $a_{-2}^{(i)}=1-a_0^{(i)}-a_{-1}^{(i)}$ using the constraint $f(0)=1$, and $i=\mathrm{I}, \mathrm{II}$.
In fig.~\ref{fig:f3} we report a comparison between a fit with function (\ref{eq:Wic}) (dashed and continuous lines for $i=\mathrm{I}, \mathrm{II}$, respectively) 
and the numerical data; the agreement is rather good for both instability types, confirming our conjecture. 

To conclude this discussion, we mention that in our calculations with a more refined grid or at $Re=0.1$ 
(see the previous section for the details about simulations) we did not observe any qualitative difference in  
the dynamical regimes occurring for different values of $\beta$ and $Wi$. 

We now consider the transition to turbulent-like flow. In the following we will present the results 
of the analysis performed for increasing $Wi$ at $\beta=1/9$. Notwithstanding some quantitative differences, 
the phenomenology holds similar in the whole range ($\beta \lesssim 0.56$) of concentrated solutions, 
including for UCM ($\beta=0$). In the case of more diluted solutions, while we observed some hints of the onset 
of irregular flow, we could not reach a fully developed regime and we cannot conclude about the emergence 
of elastic turbulence. Notice that for such large values of $\beta$, the critical Weissenberg number $Wi_c^{(\mathrm{II})}$ 
grows very rapidly, making the simulations more and more delicate.

\begin{figure}[htbp]
  \centering
  \includegraphics[width=0.38\textwidth]{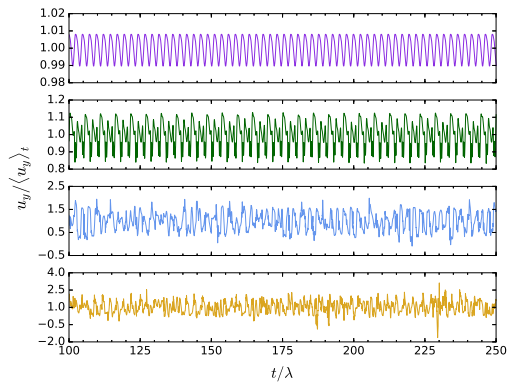}
  \caption{(Colour online) Temporal evolution (subset of the total data set, see text) of the 
  $y$-component of velocity at the outlet (probe $2$), normalised by its time average over 
  the whole time series, after the transient, 
  for $Wi=1.55,3,6,12$ (from top to bottom), $Re=0$ and $\beta=1/9$.
  }
  \label{fig:f4}
\end{figure}

Our analysis is based on the measurement of time series of the velocity components 
at two different positions marked as probe~$1$ ($\bm{x}_*^{(1)}$, entrance) and probe~$2$ ($\bm{x}_*^{(2)}$, exit) 
(see fig.~\ref{fig:f1}), 
over long durations corresponding to at least $800 \lambda$, and up to $1000 \lambda$. 
As for the experiments reported in~\cite{Sousa-18}, we choose to focus on the axial component $u_{y}(\bm{x}_*^{(2)},t)$ at the exit probe, whose  
behaviour is presented in fig.~\ref{fig:f4} for several values of $Wi$. Remark that in this figure the initial transient was removed 
and only a subset of the data record is shown.   

\begin{figure}[htbp]
  \centering
  \includegraphics[width=0.38\textwidth]{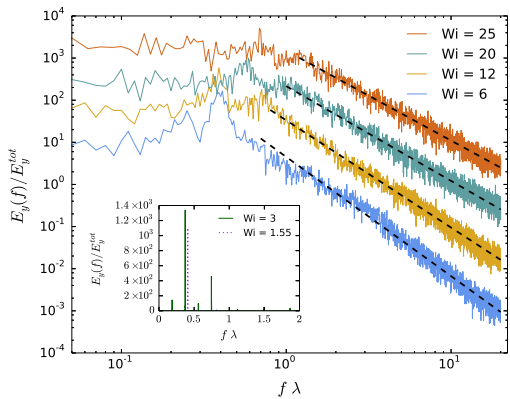}
  \caption{(Colour online) 
  Temporal spectra of fluctuations of 
  the axial velocity at the outlet $u_{y}(\bm{x}_*^{(2)},t)$, normalised by their integral $E_y^{tot}$ 
  in the elastic turbulence regime 
  for $Re=0$ and $\beta=1/9$; 
  the curves have been vertically shifted to ease readability. The dashed black curves 
  stand for 
  $E_y(f) \sim f^{-\delta}$, the fitted values of $\delta$ are 
  $\delta \simeq  (2.8, 2.5, 2.2, 2.1) \pm 0.4$ 
  for $Wi=6,12,20,25$, respectively. 
  Inset: similar spectra
  at lower elasticity. 
  For $Wi=1.55 \gtrsim Wi_c^{(\mathrm{II})}$, 
  a single frequency peak is found; at larger $Wi=3$ more discrete frequencies are present.
  }
  \label{fig:f5}
\end{figure}
\begin{figure}[htbp]
  \centering
  \includegraphics[width=0.38\textwidth]{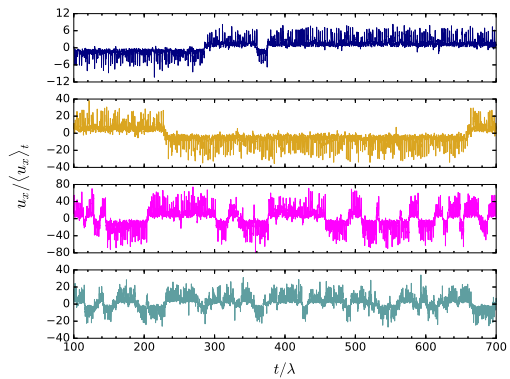}
  \caption{(Colour online) Temporal evolution 
  (subset of the total record) of 
  $u_x(\bm{x}_*^{(2)},t)$, 
  normalised by its time average over 
  the whole time series, after the transient, 
  for $Wi=11,12,15,20$ (from top to bottom), $Re=0$ and $\beta=1/9$.
  }
  \label{fig:f6}
\end{figure}

The spectra of $u_{y}(\bm{x}_*^{(2)},t)$ are shown in fig.~\ref{fig:f5}. 
All those corresponding to the developed regime are averages over 
ten 
spectra computed from consecutive subintervals 
of the velocity time series obtained for a given value of $Wi$ (after the transient). 
For $Wi \gtrsim Wi_c^{(\mathrm{II})}$, time dependency manifests in the form of regular oscillations with a single frequency close to $0.4/\lambda$ 
(see inset of fig.~\ref{fig:f5}). 
At slightly higher Weissenberg number ($Wi=3$ in fig.~\ref{fig:f4}) the flow is still periodic but it is now characterised by more discrete 
frequencies; correspondingly, the spectrum shows several distinct peaks associated with a fundamental frequency and some harmonics 
(inset of fig.~\ref{fig:f5}). The occurrence of a transitional periodic regime was also 
found in different setups~\cite{Schiamberg-06,Berti-10}. 
Above $Wi \approx 5$, the flow loses periodicity and the velocity spectra become continuous. 
Indeed, starting from $5 \lesssim Wi \lesssim 10$ they result to be quite well described by a power-law function (fig.~\ref{fig:f5}). 
When elasticity is increased in the range $Wi>10$, the faster fluctuating behaviour of the flow is accompanied by quite wide 
and irregular oscillations, over longer durations. The flow now loses its spatial asymmetry to alternatively select the outlet
in the positive/negative $y$-direction. Such a phenomenon has a strong impact on the statistics of the transversal velocity 
component $u_x(\bm{x}_*^{(2)},t)$ at the outlet (and similarly on $u_y(\bm{x}_*^{(1)},t)$ at the inlet), whose fluctuations 
are accompanied by irregular jumps between two mean values of opposite sign (see fig.~\ref{fig:f6}), 
thus complicating their analysis. A detailed 
investigation of the behaviour of such a two-state system goes beyond the scope of the present work.
  
In the turbulent-like regime ($Wi > 5$), the spectrum of velocity fluctuations displays 
a power-law behaviour $E_y(f) \sim f^{-\delta}$ 
beyond a frequency that, as in experimental studies~\cite{Sousa-18}, slightly increases with $Wi$.  
The absolute value of the exponent is found to be in the range $2 \lesssim \delta \lesssim 3$  
and shows some tendency to decrease  
at higher $Wi$; the latter feature is also detected in experiments~\cite{Sousa-18,Varshney-16}. In particular, we find 
$\delta \simeq  (2.8,2.5,2.2,2.1) \pm 0.4$ for $Wi=6,12,20,25$, respectively. 
The spectra are thus overall less steep than those previously found in experiments~\cite{Groisman-00,Sousa-18} 
and those theoretically predicted assuming homogeneity and isotropy~\cite{Fouxon-03}, 
pointing to more energetic small scales, as \textit{e.g.} the quite localised ones (fig.~\ref{fig:f2}c,d) 
stemming from intense polymer stretching, and less smooth flow. 
However, they bear an interesting similarity with those obtained in 2D numerical simulations, without artificial 
polymer-stress diffusion, of Oldroyd-B model in the presence of a cellular forcing generating distinct regions 
of strain and vorticity~\cite{Gupta-19}. 
A possible reason for the difference with the prediction of~\cite{Fouxon-03} is the 
lack of the statistical symmetries assumed by the theory in the present case. 
Indeed, our flow is neither homogeneous (due to the presence of the walls, but also of the high-strain 
region close to the centre of the setup), nor fully isotropic, as we typically observe that 
$u_y^{rms}>u_x^{rms}$ for the root-mean-square (rms) velocity components. 
Moreover, the turbulent intensity $u^{rms}/\overline{u}$, here defined as the ratio 
of the rms to the mean value of the full velocity modulus $u \equiv |\bm{u}|$ (with the overbar denoting a temporal average), 
can quite easily exceed $0.5$, and be as high as $\approx 0.8$ in conjunction with the temporal oscillations of the spatial 
asymmetry of the flow. 
Therefore, the validity of Taylor's hypothesis~\cite{Taylor-38,Frisch-95}, allowing to convert spectra from the frequency 
to the wavenumber domain, appears questionable. It might be the case that its refined version could be applied, as in~\cite{Burghelea-05}, 
but addressing this question requires further investigations. Finally, although previous numerical studies 
in two dimensions have revealed that the spectral exponent of elastic turbulence seems to be quite insensitive 
to the space dimensionality~\cite{Berti-08,Plan-17,Buel-18}, we cannot exclude that the 2D nature of our flow has an impact. 
Note, too, that values of $\delta$ for $Wi \ge 25$ should be taken with caution, as they 
may also likely depend on the length of the inlet/outlet channels.
 
\begin{figure}[htbp]
\centering
\includegraphics[width=0.38\textwidth]{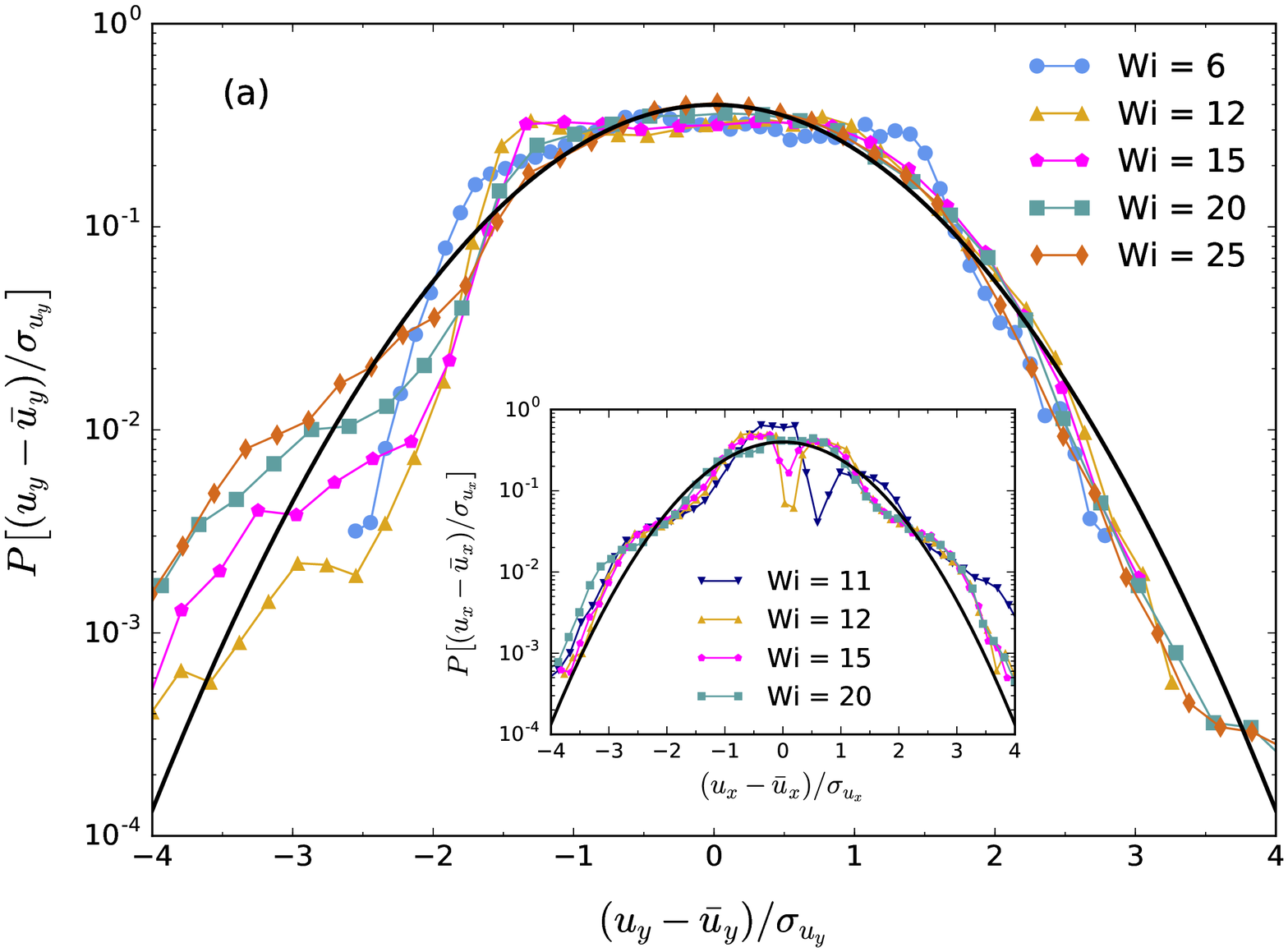}
\includegraphics[width=0.38\textwidth]{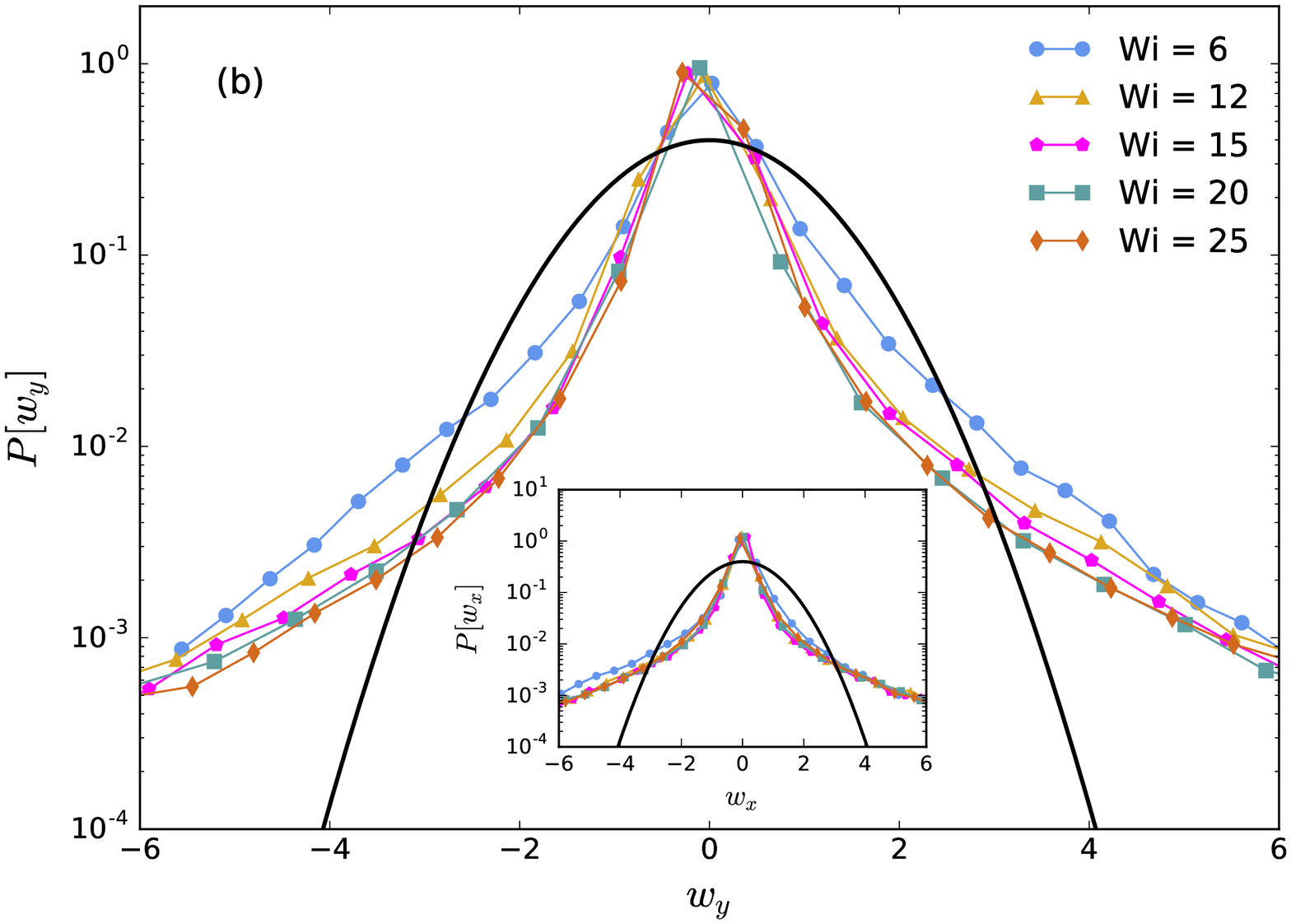}
\caption{(Colour online) Probability density functions of normalised velocity fluctuations 
$u_y'=(u_y-\overline{u_y})/\sigma_{u_y}$ (a) and temporal increments $w_y=(\partial_t u_y-\overline{\partial_t u_y})/\sigma_{\partial_t u_y}$ 
(b), where $u_y \equiv u_y(\bm{x}_*^{(2)},t)$, the overbar denotes the temporal average and $\sigma$ the standard deviation, 
for different values of $Wi$, $Re=0$ and $\beta=1/9$. 
The insets show the pdf's of the same quantities along $x-$direction (velocity and temporal-increment fluctuations, in (a) and (b), respectively). In all panels the solid black lines are standard Gaussian pdf's.}
\label{fig:f7}
\end{figure}

To further characterise the statistical properties of our elastic turbulent flows, we computed the probability density functions (pdf's) 
of the fluctuations of the velocities $u_{x,y}(\bm{x}_*^{(2)},t)$, as well as of the local accelerations $\partial_t u_{x,y}(\bm{x}_*^{(2)},t)$, 
obtained from the temporal signals at 
probe $2$. The results are presented in fig.~\ref{fig:f7}, where all variables 
are rescaled with the corresponding standard deviation $\sigma$. 
The statistics of $u_y$ fluctuations are 
not far from Gaussian, but negatively skewed for large enough $Wi$ (fig.~\ref{fig:f7}a), probably due 
to the establishment of a transversal flow component via intermittent bursts \cite{Burghelea-07}. 
Those of $u_x$ are less so (inset of fig.~\ref{fig:f7}a), instead, and show a bimodal shape for $10 \lesssim Wi \lesssim 20$, 
which reflects the importance of the flow-asymmetry alternation events in this range of elasticities. 
A qualitatively similar phenomenology is found at the entrance probe $1$, provided $x$ and $y$ indices are exchanged.    
The statistics of fluctuations of the accelerations are remarkably less dependent on the Weissenberg number (and the probe location), 
suggesting a faster (with $Wi$) onset of scaling properties at small scales.
As it can be seen in fig.~\ref{fig:f7}b, the corresponding pdf's display high tails that are indicative of non-Gaussian statistics, 
as is typical in turbulent flows and as observed in elastic turbulence experiments~\cite{Burghelea-07}.
This finding highlights the intermittent behaviour of local accelerations, likely due 
to the passage through the system of transient intense filamentary structures (fig.~\ref{fig:f2}c,d, 
but see also~\cite{Berti-10,Grilli-13,Varshney-19} about the role of elastic propagating wavy patterns).

\section*{Conclusions}
\label{sec:conclusions}
We investigated numerically the dynamics of Oldroyd-B fluids in a 2D cross-slot geometry for broad ranges 
of the Weissenberg number and the polymer concentration, focusing on the possibility to obtain elastic turbulence. 
We detected two instabilities: the first one, present only for rather concentrated solutions 
(see also~\cite{Sousa-15}), leads to steady asymmetric flow; the second one, less documented, 
manifests for all viscosity ratios $\beta < 1$ and corresponds to a supercritical Hopf bifurcation. 
By characterising the dependence of the critical Weissenberg number $Wi_c^{(\mathrm{II})}$ on the viscosity ratio, 
we found a heuristic expression that allows to quantitatively delimit the regions $Wi>Wi_c^{(\mathrm{II})}(\beta)$ 
where elastic turbulence may be excited. 

Close to the onset of the second instability, the flow of quite concentrated solutions 
displays regular oscillations in time, while at larger elasticities its dynamics 
appear more irregular. 
The frequency spectra measured in one of the outlets and far from the walls show  
distinct peaks for $Wi \gtrsim Wi_c^{(\mathrm{II})}$, while for $Wi \gtrsim 5$ they are 
well described by continuous power-law functions, of exponent $-\delta$, pointing to 
elastic turbulence. As in experiments~\cite{Sousa-18,Varshney-16}, the scaling range 
occurs beyond a frequency that moderately increases with $Wi$, and $\delta$ decreases with $Wi$. 
However, we obtain values $2<\delta<3$, somehow smaller than the experimental ones and the 
theoretical prediction for the homogeneous isotropic case~\cite{Fouxon-03}. 
While we cannot exclude an impact of the 2D nature of our flow here, and we recall the influence 
of the inlets/outlets' length on the results for $Wi \ge 25$, we remark that the 
symmetries assumed in the theory clearly do not hold for our setup. 
Similarly energetic spectra have been recently found in simulations of 2D Oldroyd-B cellular 
flows without polymer-stress diffusion~\cite{Gupta-19}. 

Further, the statistics of axial velocity components are found to be weakly non-Gaussian in the developed regime, 
while those of transversal ones also exhibit a bimodal pdf for $10<Wi<20$ due to the alternations of the spatial 
flow asymmetry occurring in this range of $Wi$. 
The pdf's of both components of the local accelerations, 
instead, present high non-Gaussian tails indicative of intermittency. 
Such a phenomenology agrees with that observed in experiments~(see \textit{e.g.} \cite{Burghelea-07}).  

In summary, we reproduced the different dynamical regimes experimentally observed in cross-slot devices, 
and we obtained turbulent-like states bearing good statistical resemblance with elastic turbulence. 
The quantitative differences highlighted call for further theoretical and numerical developments. 
In the future it would be interesting to explore such dynamics in three-dimensional flows.

\begin{acknowledgments}
D. O. Canossi acknowledges financial support from a PhD grant funded by the Brazilian agency CNPq 
(\textit{Conselho Nacional de Desenvolvimento Cient\'ifico e Tecnol\'ogico}). We are grateful to N. Ouarzazi 
and E. Calzavarini for interesting discussions about instabilities and turbulent-like features, respectively.
\end{acknowledgments}

\bibliography{references}

\end{document}